\documentclass{article}
\usepackage{spconf,amsmath,graphicx}
\usepackage{subcaption}

\title{Single Channel Audio Source Separation using Convolutional Denoising Autoencoders}
\name{Emad M. Grais and Mark D. Plumbley}
\address{Centre for Vision, Speech and Signal Processing, University of Surrey, Guildford, UK.}
\begin{document}
\ninept
\maketitle
\begin{abstract}
Deep learning techniques have been used recently to tackle the audio source separation problem. In this work, we propose to use deep fully convolutional denoising autoencoders (CDAEs) for monaural audio source separation. We use as many CDAEs as the number of sources to be separated from the mixed signal. Each CDAE is trained to separate one source and treats the other sources as background noise. The main idea is to allow each CDAE to learn suitable spectral-temporal filters and features to its corresponding source. Our experimental results show that CDAEs perform source separation slightly better than the deep feedforward neural networks (FNNs) even with fewer parameters than FNNs.
\end{abstract}
\section{Introduction}
\label{sec:intro}
Different types of deep neural networks (DNNs) have been used to tackle the single channel source separation (SCSS) problem for audio mixtures \cite{Stefan:17:imssbdnntdanb,Grais:17:descassdnn, Kim:15:adaatsltm, chandna:17:massudcnn}. The denoising autoencoder (DAE) is a special type of fully connected feedforward neural networks that takes noisy input signals and outputs their denoised version \cite{Xie:12:ididnn,Vincent:10:sdalurdnldc}. DAEs are common in deep learning, they are used to learn noise robust low-dimensional features even when the inputs are perturbed with some noise \cite{Pascal:08:ecrfda,HinSal:06:rddnn}. DAEs have been used for SCSS where the inputs of the DAE are the spectral frames of the mixed signal and the outputs are the spectral frames of the target source \cite{Kim:15:adaatsltm, Paris:17:nnanmf}. The fully connected DAEs with frame-wise inputs and outputs do not capture the 2D (spectral-temporal) structures of the spectrogram of the input and output signals. Since DAEs are fully connected networks, they usually have many parameters to be optimized.

For their ability in extracting robust spectral-temporal structures of different audio signals \cite{Honglak:09:uflaccdbn}, convolutional neural networks (CNN) have been used successfully to learn useful features in many audio processing applications such as: speech recognition \cite{Yanmin:16:vdcnnnrsr}, speech enhancement \cite{Szu:16:snrcnnmse}, audio tagging \cite{Yong:17:cgrnnisfat}, and many music related applications \cite{Yoonchang:17:dcnnpirpm,Keunwoo:16:crnnmc,Filip:16:fcdammcr}. Convolutional denoising autoencoders (CDAEs) are also a special type of CNNs that can be used to discover robust localized low-dimensional patterns that repeat themselves over the input \cite{Jonathan:11:scaehfe,Bo:17:scdaefr}. CDAEs differ from conventional DAEs as their parameters (weights) are shared, which makes the CDAEs have fewer parameters than DAEs. The ability of CDAEs to extract repeating patterns in the input makes them suitable to be used to extract speech signals from background noise and music signals for speech enhancement and recognition \cite{se:16:fcnnse,Mengyuan:16:mrcdasr}. 

Motivated by the aforementioned successes of using neural networks with convolutional layers in a variety of audio signal processing applications, we propose in this paper to use deep fully convolutional denoising autoencoders, where all the layers of the CDAEs are composed of convolutional units, for single channel source separation (SCSS). The main idea in this paper is to train a CDAE to extract one target source from the mixture and treats the other sources as background noise that needs to be suppressed. This means we need as many CDAEs as the number of sources that need to be separated from the mixed signal. This is a very challenging task because each CDAE has to deal with highly nonstationary background signals/noise. Each CDAE sees the magnitude spectrograms as 2D segments which helps in learning the spectral and temporal information for the audio signals. From the ability of CDAEs in learning noise robust features, in this work, we train each CDAE to learn unique spectral-temporal patterns for its corresponding target source. Each trained CDAE is then used to extract/separate the related patterns of its corresponding target source from the mixed signal.

This paper is organized as follows: Section \ref{fcda} shows a brief introduction about CDAEs. The proposed approach of using CDAEs for SCSS is presented in Section \ref{overall}. The experiments and discussions are shown in Section \ref{sec:exp}. 

\begin{figure*}[h]
 \includegraphics[width=1\linewidth,height=4.5cm]{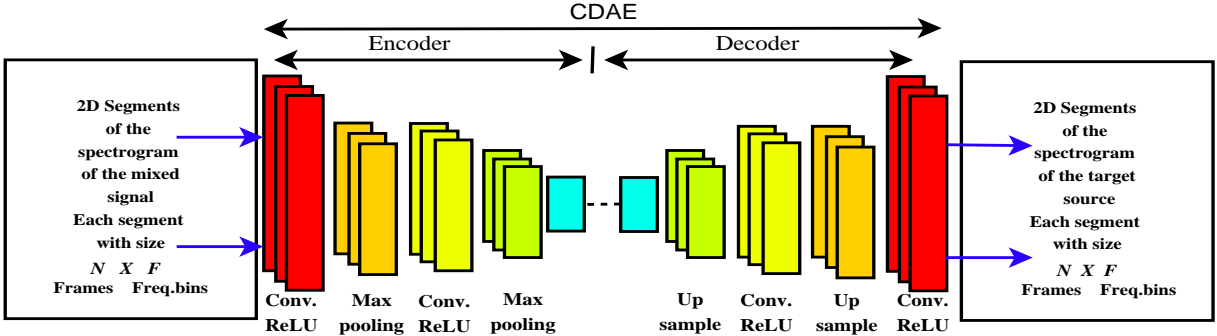}
\caption{\label{fig:cdae1}{The overview of the proposed structure of a convolutional denoising auto-encoder (CDAE) that separates one target source from the mixed signal. ``Conv.'' denotes a 2D convolutional layer, ReLU denotes a rectified linear unit as an activation function. We use a CDAE for each source.}}
\end{figure*}

\section{Fully convolutional denoising autoencoders}
\label{fcda}
Fully convolutional autoencoders (CAEs) \cite{Jonathan:11:scaehfe,Bo:17:scdaefr} are composed of two main parts, the encoder part and the decoder part. The encoder part maps the input data into low dimensional features. The decoder part reconstructs the input data from the low dimensional features. Convolutional denoising autoencoders (CDAEs) are similar to CAEs but CDAEs are trained from corrupted input signals and the encoder is used to extract noise robust features that the decoder can use to reconstruct a cleaned-up version of the input data \cite{se:16:fcnnse,Mengyuan:16:mrcdasr}. The encoder part in CDAEs is composed of repetitions of a convolutional layer, an activation layer, and a pooling layer as shown in Fig. \ref{fig:cdae1}. The convolutional layers consist of a set of filters that extract features from their input layers, the activation layer in this work is the rectified linear unit that imposes nonlinearity to the feature maps. The pooling in this work is chosen to be max-pooling \cite{Dominik:10:epocaor}. The max-pooling does the down-sampling of the latent representation by a constant factor taking the maximum value within a certain scope of the mapping space and generates a new mapping space with a reduced dimension. 
The final goal of the  encoder part is to extract noise robust low dimensional features from the input data. The max-pooling is the layer that reduces the dimensionality of the mapping space. The decoder part consists of repetitions of a convolutional layer, an activation layer, and an up-sampling layer. The up-sampling layer does the up-sampling on the feature maps of the previous layer and generates new ones with high dimension. In this work, the data are 2D signals (magnitude spectrograms). The filtering, pooling, and up-sampling are all 2D operators. 
%

\section{The proposed approach of using CDAEs for SCSS}
\label{overall}
Given a mixture of $I$ sources as $y(t) = \sum_{i=1}^I s_i(t)$, the aim of audio SCSS is to estimate the sources $s_i(t), \ \forall{i}$, from the mixed signal $y(t)$ \cite{emad:12:avsrwbmscss,emad:12:stpsnmfscss}. We work here in the short-time Fourier transform (STFT) domain. Given the STFT of the mixed signal $y(t)$, the main goal is to estimate the STFT of each source in the mixture.

In this work we propose to use CDAEs for source separation. We propose to use as many CDAEs as the number of sources to be separated from the mixed signal. Each CDAE sees the mixed signal as a combination of its target source and background noise. The main aim of each CDAE is to estimate a clean signal for its corresponding source from the other background sources that exist in the mixed signal. This is a challenging task for each CDAE since each CDAE deals with highly nonstationary background noise (other sources in the mixture). Each CDAE is trained to map the magnitude spectrogram of the mixture into the magnitude spectrogram of its corresponding target source. Each CDAE in this work is a fully 2D convolutional deep neural network without any fully connected layer, which keeps the number of parameters to be optimized for each CDAE very small. Also using fully 2D convolutional layers allows neat 2D spectral-temporal representations for the data through all the layers in the network while considering the spectral-temporal representations in the case of using fully connected layers requires stacking multiple consecutive frames to form very long feature vectors. The inputs and outputs of the CDAEs are 2D-segments from the magnitude spectrograms of the mixed and target signals respectively. Therefore, the CDAEs span multiple time frames to capture the spectral-temporal characteristics of each source. The number of frames that each input segment has is $N$ and the number of frequency bins is $F$. In this work, $F$ is the dimension of the whole spectral frame.  
%
%
\subsection{Training the CDAEs for source separation}
\label{sec:train}
Let's assume we have training data for the mixed signals and their corresponding clean/target sources. Let $\mathbf{Y}_{\mbox{tr}}$ be the magnitude spectrogram of the mixed signal and $\mathbf{S}_i$ be the magnitude spectrogram of the clean source $i$. The subscript ``tr'' denotes the training data. The CDAE that separates source $i$ from the mixture is trained to minimize the following cost function:
\begin{equation}
\label{cost_mask}
C_i =\sum_{n,f}\left( \mathbf{Z}_i\left(n,f\right) - \mathbf{S}_i\left(n,f\right) \right)^2
\end{equation}
where $\mathbf{Z}_i$ is the actual output of the last layer of the CDAE of source $i$, $\mathbf{S}_i$ is the reference clean output signal for source $i$, $n$, and $f$ are the time and frequency indices respectively. The input of all the CDAEs is the magnitude spectrogram $\mathbf{Y}_{\mbox{tr}}$ of the mixed signal. 

Note that the input and output instants of the CDAEs are 2D-segments from the spectrograms of the mixed and target signals respectively. Each segment is composed of $N$ consecutive spectral frames taken from the magnitude spectrograms. This allows the CDAEs to learn unique spectral-temporal patterns for each source. 
%
%
\subsection{Testing the CDAEs for source separation}
\label{sec:test} 
Given the trained CDAEs, the magnitude spectrogram $\mathbf{Y}$ of the mixed signal is passed through all the trained CDAEs. The output of the CDAE of source $i$ is the estimate $\mathbf{\tilde{S}}_i$ of the spectrogram of source $i$. 

\section{Experiments and Discussion}
\label{sec:exp}
We applied our proposed single channel source separation (SCSS) using CDAEs approach to separate audio sources from a group of songs from the SiSEC-2015-MUS-task dataset \cite{ono:15:tsisec}. The dataset has 100 stereo songs with different genres and instrumentations. To use the data for the proposed SCSS approach, we converted the stereo songs into mono by computing the average of the two channels for all songs and sources in the data set. Each song is a mixture of vocals, bass, drums, and other musical instruments. We used our proposed algorithm to separate each song into vocals, bass, drums and other instruments. The other instruments (\textit{other} for short) here (sources that are not vocal, bass, or drums) are treated as one source. We trained four CDAEs for the four sources (vocals, bass, drums, and other). 

The first 50 songs were used as training and validation datasets to train all the networks for separation, and the last 50 songs were used for testing. The data was sampled at 44.1kHz. The magnitude spectrograms for the data were calculated using the STFT, a Hanning window with 2048 points length and overlap interval of 512 was used and the FFT was taken at 2048 points, the first 1025 FFT points only were used as features since the conjugate of the remaining 1024 points are involved in the first points.

For the input and output data for the CDAEs, we chose the number of spectral frames in each 2D-segment to be 15 frames. This means the dimension of each input and output instant for each CDAE is 15 (time frames) $\times$ 1025 (frequency bins). Thus, each input and output instant (the 2D-segments from the spectrograms) spans around 370 msec of the waveforms of the data.  
\begin{table}
\scalebox{0.95}
{
\begin{tabular}{ |p{3cm}||p{2.5cm}|p{2cm}|}
 \hline
 \multicolumn{3}{|c|}{CDAE model summary} \\
 \hline
 \multicolumn{3}{|c|}{The input data with size 15 frames and 1025 frequency bins} \\
 \hline
 Layer (type)         & Number of filters & output shape \\
  \hline
 Convolution2D(3,3)   & 12    & (15, 1025) \\
  Max-Pooling2D(3,5)    &        & (5, 205) \\
  \hline 
 Convolution2D        & 20     & (5, 205) \\
  Max-Pooling2D(1,5)    &        & (5, 41) \\
 \hline 
 Convolution2D        & 30     & (5, 41) \\
  \hline 
 Convolution2D        & 40     & (5, 41) \\
  \hline 
 Convolution2D        & 30     & (5, 41) \\
  \hline 
 Convolution2D        & 20      & (5, 41) \\
  Up-Sampling2D(1,5)   &        & (5, 205) \\
  \hline 
Convolution2D(3,3)    & 12      & (5, 205) \\
Up-Sampling2D(3,5)     &       & (15, 1025) \\
\hline 
 Convolution2D        & 1      & (15, 1025) \\
  \hline
 \multicolumn{3}{|c|}{The output data with size 15 frames and 1025 frequency bins} \\
 \multicolumn{3}{|c|}{Total number of parameters: 37,101}\\
 \hline
\end{tabular}
}
\caption{{The detail structure of each CDAE. The output shape is shown as (time-frame , frequency). ``Convolution2D(3,3)'' denotes 2D convolutional layer with filter size 3$\times$3. ``Max-Pooling2D(3,5)'' denotes down-sampling by 3 in the time-frame direction and by 5 in the frequency direction. ``Up-Sampling2D(3,5)'' denotes up-sampling by 3 in the time-frame direction and by 5 in the frequency direction.}}
\label{table:cdae} 
\end{table}

For the CDAEs, each CDAE has seven hidden 2D convolutional layers with rectified linear unit (ReLU) as an activation function. The details of the dimensions of the input, convolutional, max-pooling, up-sampling and output layers of each CDAE are shown in Table \ref{table:cdae}. The dimensions are shown as time-frames $\times$ frequency. The size of each filter is 3$\times$3 as in \cite{Yanmin:16:vdcnnnrsr,Filip:16:fcdammcr}. ``Max-Pooling2D(3,5)'' in Table \ref{table:cdae}, denotes down-sampling the feature maps by 3 in the time-frame direction and by 5 in the frequency direction of the 2D feature maps. ``Up-Sampling2D(3,5)'' in Table \ref{table:cdae}, denotes up-sampling the feature maps by 3 in the time-frame direction and by 5 in the frequency direction. The output layer is also a convolutional layer with ReLU that its size was adjusted to match the size of each 2D output segment (15$\times$1025) of the spectrogram. As can be seen from Table \ref{table:cdae}, each CDAE has 37,101 parameters. 

\begin{table}[]
\centering
\caption{The number of parameters for each FNN and CDAE.}
\label{table:parmtrs}
\begin{tabular}{ll}
\hline
FNN & 4,206,600  \\
\hline
CDAE & 37,101  \\
\hline
\end{tabular}
\end{table}

We compared our proposed SCSS using CDAEs approach for SCSS with using deep fully connected feedforward neural networks (FNN) for SCSS approach. Four FNNs were used and each FNN was used to separate one source. Each FNN has three hidden layers with ReLU as activation functions. Each hidden layer has 1025 nodes. The parameters of the networks are tuned based on our previous work on the same dataset \cite{Emad:16:scassdnne, grais:16:cmescassdnn}. As shown in Table \ref{table:parmtrs}, the number of parameters in each FNN is 4,206,600 parameters which is greater than 100 times the number of parameters in each CDAE.

\begin{figure}
\centering
   \begin{subfigure}[b]{0.43\textwidth}
   \includegraphics[width=0.99\linewidth]{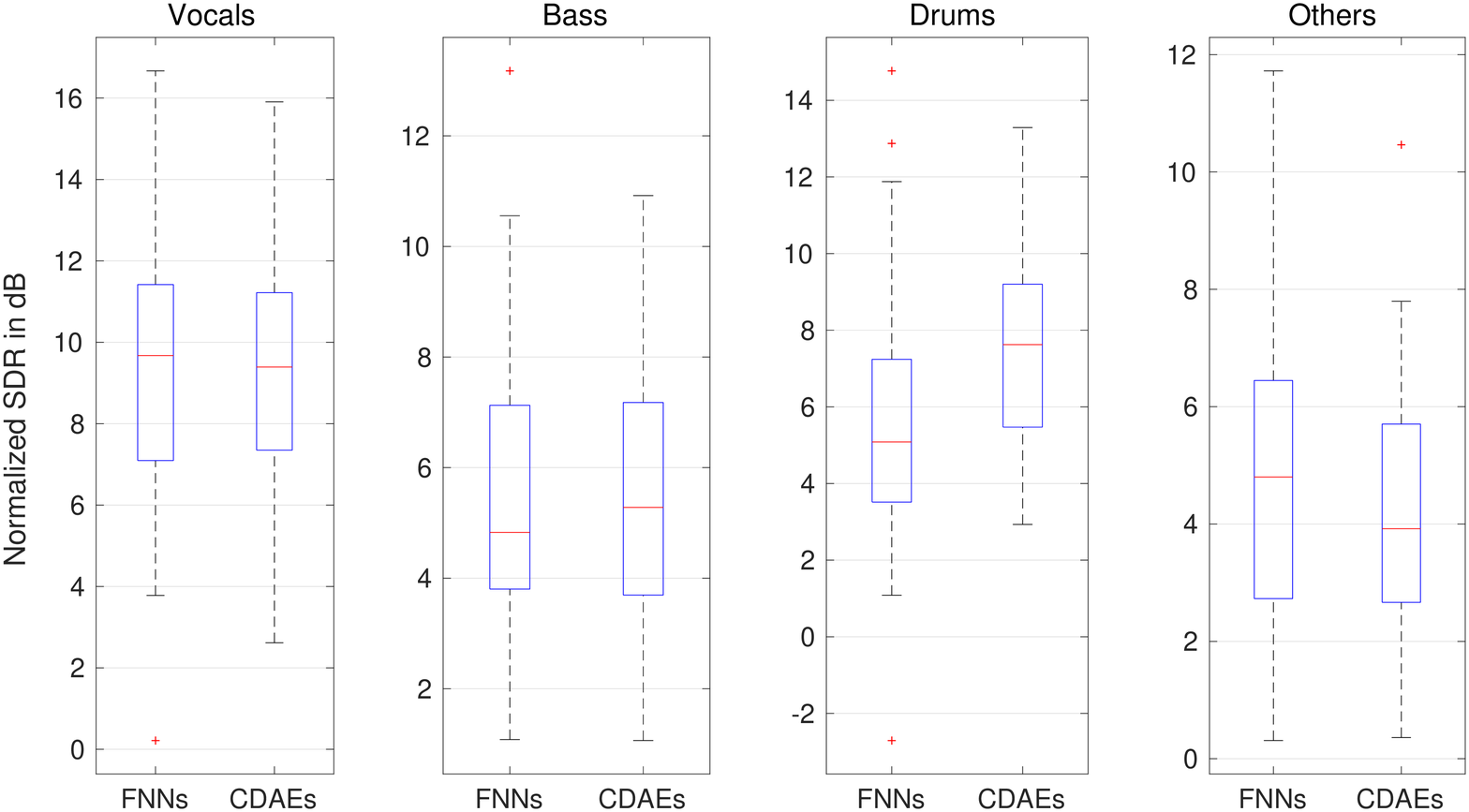} 
   \caption{Normalized SDR in dB}\hfill \hfill\hfill\hfill\hfill\hfill
   \label{fig:dnn1} 
\end{subfigure}

\begin{subfigure}[b]{0.43\textwidth}
   \includegraphics[width=0.99\linewidth]{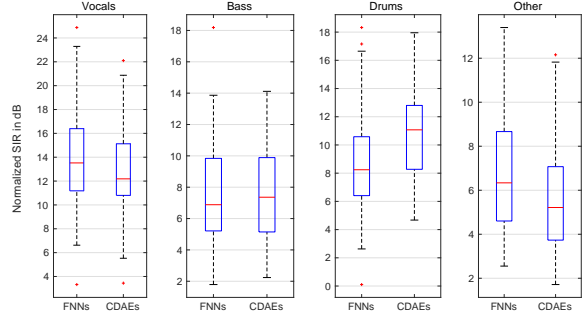} 
   \caption{Normalized SIR in dB}\hfill\hfill\hfill\hfill\hfill\hfill 
   \label{fig:dnn2}
\end{subfigure}

\begin{subfigure}[b]{0.43\textwidth}
   \includegraphics[width=0.99\linewidth]{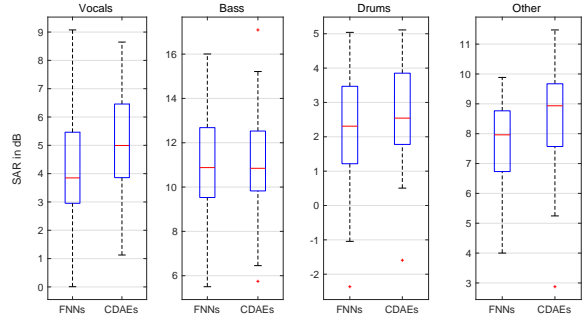}
   \caption{SAR in dB}\hfill\hfill
   \label{fig:dnn3} 
\end{subfigure}

\caption[]{(a) The normalized SDR, (b) the normalized SIR, and (c) SAR values in dB of using deep fully connected feedforward neural networks (FNNs) and the proposed method of using deep fully convolutional denoising autoencoders (CDAEs) for source separation.}
\label{fig:dnn_dnn} 
\end{figure}

The parameters for all the networks were initialized randomly. All the networks were trained using backpropagation with Nesterovs accelerated gradient \cite{Nesterov:83:mscppcr} as recommended by \cite{icml2013_sutskever13} with parameters: $\beta_1=0.9$, $\beta_2=0.999$, $\epsilon=1e-08$, schedule-decay$=0.004$, batch size 100, and a learning rate starts with $0.002$ and reduced by a factor of 10 when the values of the cost function do not decrease on the validation set for 3 consecutive epochs. The maximum number of epochs is 100. We implemented our proposed algorithm using Keras based on Theano \cite{chollet2015keras,Bastien:12:theano1}.

Since in this work we separate all four sources from the mixed signals, it is usually preferred to separate the sources from the mixed signal by building spectral masks that scale the mixed signal according to the contribution of each source in the mixed signal \cite{arie:16:masswdnn,Huang:14:svsmrdrnn}. The masks make sure that the sum of the estimated sources adds up to the mixed signal \cite{erdogan:10:sbsmsusacp,grais:14:ssrnmfmmse}. Here we used the output spectrograms $\mathbf{\tilde{S}}_i,\ \forall i$ of the networks to build spectral masks as follows:
\begin{equation}
\label{mask}
\footnotesize{
\mathbf{M}_i (n,f)= \frac{\mathbf{\tilde{S}}_i(n,f)}{\sum_j^I \mathbf{\tilde{S}}_j(n,f)}, \ \ \forall i.
}
\end{equation}
The final estimate for source $i$ can be found as
\begin{equation}
\label{mask}
\footnotesize{
\mathbf{\hat{S}}_i(n,f) =  \mathbf{M}_i (n,f) \times \mathbf{Y}(n,f)
}
\end{equation}
where $\mathbf{Y}(n,f)$ is the magnitude spectrogram of the mixed signal at frame $n$ and frequency $f$. The time domain estimate for source $\hat s_i(t)$ is computed using the inverse STFT of $\hat {\mathbf{S}}_i$ with the phase angle of the STFT of the mixed signal.

The quality of the separated sources was measured using the signal to distortion ratio (SDR), signal to interference ratio (SIR), and signal to artefact ratio (SAR) \cite{vincent:06:pmi}. SIR indicates how well the sources are separated based on the remaining interference between the sources after separation. SAR indicates the artefacts caused by the separation algorithm in the estimated separated sources. SDR measures how distorted the separated sources are. The SDR values are usually considered as the overall performance evaluation for any source separation approach \cite{vincent:06:pmi}. Achieving high SDR, SIR, and SAR indicates good separation performance. 

Figs. \ref{fig:dnn1} and \ref{fig:dnn2} show the box-plots of the normalized SDR and SIR values respectively. The normalization was done by subtracting the SDR and SIR values of the mixed signal from their corresponding values of the estimated sources \cite{alexey:07:aobmfscssaiatvmsips}. Fig. \ref{fig:dnn3} shows the box-plots of the SAR values of the separated sources using FNNs and CDAEs. We show the SAR values without normalization because SAR for the mixed signal is usually very high. 

As can be seen from Fig. \ref{fig:dnn_dnn}, CDAEs with few parameters improve the quality of the sources by achieving positive normalized SDR and SIR values and high SAR values for the separated sources. We can also see that CDAEs work significantly better than FNNs for drums, which is considered as a difficult source to be separated \cite{Stefan:17:imssbdnntdanb,arie:16:masswdnn}. For the SDR and SIR, the performance of FNNs and CDAEs is almost the same for the vocals and bass sources. For SAR, the performance of CDAEs and FNNs is the same for bass, but CDAEs perform better than FNNs in the remaining sources. In general, it is not easy to have a fair comparison between the two methods since FNNs have more parameters than CDAEs and CDAEs were applied on spectral-temporal segments (without overlapping) while the FNNs were applied on individual spectral frames. 

\section{Conclusions}
In this work we proposed a new approach for single channel source separation (SCSS) of audio mixtures. The new approach is based on using deep fully convolutional denoising autoencoders (CDAEs). We used as many CDAEs as the number of sources to be separated from the mixed signal. Each CDAE learns unique patterns for each source and uses this information to separate the related components of each source from the mixed signal. The experimental results indicate that using CDAEs for SCSS is a promising approach and with very few parameters can achieve competitive results with the feedforward neural networks. 

In our future work, we will investigate the effect of changing the parameters in the CDAE including: the size of the filters, the number of filters in each layer, the max-pooling and up-sampling ratios, the number of frames in each input/output segment, and the number of layers on the quality of the separated sources. We will also investigate the possibility of using CDAEs for the multi-channel audio source separation problem using multi-channel convolutional neural networks.     

\section{Acknowledgements}
 This work is supported by grants EP/L027119/1 and EP/L027119/2 from the UK Engineering and Physical Sciences Research Council (EPSRC).
%
%
\bibliographystyle{IEEEbib}
\bibliography{refs}

\begin{thebibliography}{10}

\bibitem{Stefan:17:imssbdnntdanb}
S.~Uhlich, M.~Porcu, F.~Giron, M.~Enenkl, T.~Kemp, N.~Takahashi, and
  Y.~Mitsufuji,
\newblock ``{Improving} music source separation based on deep neural networks
  through data augmentation and network blending,''
\newblock in {\em Proc. ICASSP}, 2017.

\bibitem{Grais:17:descassdnn}
E.~M. Grais, G.~Roma, A.~J.R. Simpson, and M.~D. Plumbley,
\newblock ``{Discriminative} enhancement for single channel audio source
  separation using deep neural networks,''
\newblock in {\em Proc. LVA/ICA}, 2017, pp. 236--246.

\bibitem{Kim:15:adaatsltm}
M.~Kim and P.~Smaragdis,
\newblock ``{Adaptive} denoising autoencoders: A fine-tuning scheme to learn
  from test mixtures,''
\newblock in {\em Proc. LVA/ICA}, 2015, pp. 100--107.

\bibitem{chandna:17:massudcnn}
P.~Chandna, M.~Miron, J.~Janer, and E.~Gomez,
\newblock ``{Monoaural} audio source separation using deep convolutional neural
  networks,''
\newblock in {\em Proc. LVA/ICA}, 2017, pp. 258--266.

\bibitem{Xie:12:ididnn}
J.~Xie, L.~Xu, and E.~Chen,
\newblock ``{Image} denoising and inpainting with deep neural networks,''
\newblock in {\em Advances in NIPS}, 2012.

\bibitem{Vincent:10:sdalurdnldc}
P.~Vincent, H.~Larochelle, I.~Lajoie, Y.~Bengio, and P.~A. Manzagol,
\newblock ``{Stacked Denoising Autoencoders:} learning useful representations
  in a deep network with a local denoising criterion,''
\newblock {\em Journal of Machine Learning Research}, vol. 11, pp. 3371 --
  3408, 2010.

\bibitem{Pascal:08:ecrfda}
P.~Vincent, H.~Larochelle, Y.~Bengio, and P.~A. Manzagol,
\newblock ``{Extracting} and composing robust features with denoising
  autoencoders,''
\newblock in {\em Proc. ICML}, 2008, pp. 1096--1103.

\bibitem{HinSal:06:rddnn}
G.~Hinton and R.~Salakhutdinov,
\newblock ``{Reducing} the dimensionality of data with neural networks,''
\newblock {\em Science}, vol. 313, no. 5786, pp. 504 -- 507, 2006.

\bibitem{Paris:17:nnanmf}
P.~Smaragdis and S.~Venkataramani,
\newblock ``{A} neural network alternative to non-negative audio models,''
\newblock in {\em Proc. ICASSP}, 2017.

\bibitem{Honglak:09:uflaccdbn}
H.~Lee, P.~Pham, Y.~Largman, and A.~Y. Ng,
\newblock ``{Unsupervised} feature learning for audio classification using
  convolutional deep belief networks,''
\newblock in {\em Advances in NIPS}, 2009, pp. 1096--1104.

\bibitem{Yanmin:16:vdcnnnrsr}
Y.~Qian, M.~Bi, T.~Tan, and K.~Yu,
\newblock ``{Very} deep convolutional neural networks for noise robust speech
  recognition,''
\newblock {\em IEEE Trans. on Audio, Speech, and Language Processing}, vol. 24,
  no. 12, pp. 2263--2276, 2016.

\bibitem{Szu:16:snrcnnmse}
S.~W. Fu, Y.~Tsao, and X.~Lu,
\newblock ``{SNR-Aware} convolutional neural network modeling for speech
  enhancement,''
\newblock in {\em Proc. InterSpeech}, 2016.

\bibitem{Yong:17:cgrnnisfat}
Y.~Xu, Q.~Kong, Q.~Huang, W.~Wang, and M.~D. Plumbley,
\newblock ``{Convolutional} gated recurrent neural network incorporating
  spatial features for audio tagging,''
\newblock in {\em Proc. IJCNN}, 2017.

\bibitem{Yoonchang:17:dcnnpirpm}
Y.~Han, J.~Kim, and K.~Lee,
\newblock ``{Deep} convolutional neural networks for predominant instrument
  recognition in polyphonic music,''
\newblock {\em IEEE Trans. on Audio, Speech, and Language Processing}, vol. 25,
  no. 1, pp. 208--221, 2017.

\bibitem{Keunwoo:16:crnnmc}
K.~Choi, G.~Fazekas, M.~Sandler, and K.~Cho,
\newblock ``{Convolutional} recurrent neural networks for music
  classification,''
\newblock in {\em arXiv:1609.04243}, 2016.

\bibitem{Filip:16:fcdammcr}
F.~Korzeniowski and G.~Widmer,
\newblock ``{A} fully convolutional deep auditory model for musical chord
  recognition,''
\newblock in {\em Proc. Workshop on Machine Learning for Signal Processing},
  2016.

\bibitem{Jonathan:11:scaehfe}
J.~Masci, U.~Meier, D.~Ciresan, and J.~Schmidhuber,
\newblock ``{Stacked} convolutional auto-encoders for hierarchical feature
  extraction,''
\newblock in {\em Advances in NIPS}, 2011.

\bibitem{Bo:17:scdaefr}
B.~Du, W.~Xiong, J.~Wu, L.~Zhang, L.~Zhang, and D.~Tao,
\newblock ``{Stacked} convolutional denoising auto-encoders for feature
  representation,''
\newblock {\em IEEE Trans. on Cybernetics}, vol. 47, no. 4, pp. 1017--1027,
  2017.

\bibitem{se:16:fcnnse}
S.~R. Park and J.~W. Lee,
\newblock ``{A} fully convolutional neural network for speech enhancement,''
\newblock in {\em arXiv preprint arXiv:1609.07132}, 2016.

\bibitem{Mengyuan:16:mrcdasr}
M.~Zhao, D.~Wang, Z.~Zhang, and X.~Zhang,
\newblock ``{Music} removal by convolutional denoising autoencoder in speech
  recognition,''
\newblock in {\em In proc. APSIPA}, 2016.

\bibitem{Dominik:10:epocaor}
D.~Scherer, A.~Muller, and S.~Behnke,
\newblock ``{Evaluation} of pooling operations in convolutional architectures
  for object recognition,''
\newblock in {\em Advances in NIPS}, 2010.

\bibitem{emad:12:avsrwbmscss}
E.~M. Grais, I.~S. Topkaya, and H.~Erdogan,
\newblock ``{Audio-Visual} speech recognition with background music using
  single-channel source separation,''
\newblock in {\em Proc. SIU}, 2012.

\bibitem{emad:12:stpsnmfscss}
E.~M. Grais and H.~Erdogan,
\newblock ``{Spectro-temporal} post-smoothing in {NMF} based single-channel
  source separation,''
\newblock in {\em Proc. EUSIPCO}, 2012.

\bibitem{ono:15:tsisec}
N.~Ono, Z.~Rafii, D.~Kitamura, N.~Ito, and A.~Liutkus,
\newblock ``{The 2015 signal separation evaluation campaign},''
\newblock in {\em Proc. LVA/ICA}, 2015, pp. 387--395.

\bibitem{Emad:16:scassdnne}
E.~M. Grais, G.~Roma, A.~J.~R. Simpson, and M.~D. Plumbley,
\newblock ``{Single} channel audio source separation using deep neural network
  ensembles,''
\newblock in {\em Proc. 140th Audio Engineering Society Convention}, 2016.

\bibitem{grais:16:cmescassdnn}
E.~M. Grais, G.~Roma, A.~J.~R. Simpson, and M.~D Plumbley,
\newblock ``{Combining} mask estimates for single channel audio source
  separation using deep neural networks,''
\newblock in {\em Prec. InterSpeech}, 2016.

\bibitem{Nesterov:83:mscppcr}
Y.~Nesterov,
\newblock ``{A method} of solving a convex programming problem with convergence
  rate o(1/sqr(k)),''
\newblock {\em Soviet Mathematics Doklady}, vol. 27, no. 2, pp. 372--376, 1983.

\bibitem{icml2013_sutskever13}
I.~Sutskever, J.~Martens, G.~Dahl, and G.~Hinton,
\newblock ``{On} the importance of initialization and momentum in deep
  learning,''
\newblock in {\em Proc. ICML}, May 2013, vol.~28, pp. 1139--1147.

\bibitem{chollet2015keras}
F.~Chollet,
\newblock ``Keras, https://github.com/fchollet/keras,'' 2015.

\bibitem{Bastien:12:theano1}
F.~Bastien, P.~Lamblin, R.~Pascanu, J.~Bergstra, I.~Goodfellow, A.~Bergeron,
  N.~Bouchard, D.~W. F., and Y.~Bengio,
\newblock ``{Theano: new features and speed improvements},''
\newblock in {\em Deep Learning and Unsupervised Feature Learning NIPS
  Workshop}, 2012.

\bibitem{arie:16:masswdnn}
A.~A. Nugraha, A.~Liutkus, and E.~Vincent,
\newblock ``{Multichannel} audio source separation with deep neural networks,''
\newblock {\em IEEE/ACM Trans. on Audio, Speech, and Language Processing}, vol.
  24, no. 9, pp. 1652--1664, 2016.

\bibitem{Huang:14:svsmrdrnn}
P.~S. Huang, M.~Kim, M.~H. J., and P.~Smaragdis,
\newblock ``{Singing-Voice} separation from monaural recordings using deep
  recurrent neural networks,''
\newblock in {\em Proc. ISMIR}, 2014, pp. 477--482.

\bibitem{erdogan:10:sbsmsusacp}
H.~Erdogan and E.~M. Grais,
\newblock ``{Semi-blind} speech-music separation using sparsity and continuity
  priors,''
\newblock in {\em ICPR}, 2010.

\bibitem{grais:14:ssrnmfmmse}
E.~M. Grais and H.~Erdogan,
\newblock ``{Source} separation using regularized {NMF} with {MMSE} estimates
  under {GMM} priors with online learning for the uncertainties,''
\newblock {\em Digital Signal Processing}, vol. 29, pp. 20--34, 2014.

\bibitem{vincent:06:pmi}
E.~Vincent, R.~Gribonval, and C.~Fevotte,
\newblock ``{Performance} measurement in blind audio source separation,''
\newblock {\em IEEE Trans. on Audio, Speech, and Language Processing}, vol. 14,
  no. 4, pp. 1462--69, July 2006.

\bibitem{alexey:07:aobmfscssaiatvmsips}
A.~Ozerov, P.~Philippe, F.~Bimbot, and R.~Gribonval,
\newblock ``{Adaptation of Bayesian} models for single-channel source
  separation and its application to voice/music separation in popular songs,''
\newblock {\em IEEE Trans. of Audio, Speech, and Language Processing}, vol. 15,
  pp. 1564--1578, 2007.

\end{thebibliography}

\end{document}